\documentclass{article}
\usepackage{graphicx} 
\graphicspath{ {images/} }
\usepackage{booktabs,tabularx}
\usepackage{float}
\usepackage{svg}
\usepackage{caption}
\usepackage{indentfirst}
\usepackage{makecell}
\usepackage{array}
\usepackage{changepage}
\usepackage{adjustbox}
\usepackage{multirow}
\usepackage{comment}
\usepackage{amsfonts}
\usepackage{verbatim} 
\usepackage{amsmath} 
\usepackage{amssymb} 
\usepackage{fullpage} 
\usepackage{paralist} 
\usepackage{listings}
\usepackage{subfig}  
\usepackage{enumitem} 
\usepackage{siunitx} 
\usepackage{tikz,bm} 
\usepackage{circuitikz}
\usepackage{tikz}
\usepackage{mathtools}
\usepackage[outline]{contour}
\usepackage{upgreek}
\usepackage{empheq}
\usepackage{caption}
\DeclarePairedDelimiter\bra{\langle}{\rvert}
\DeclarePairedDelimiter\ket{\lvert}{\rangle}
\DeclarePairedDelimiterX\braket[2]{\langle}{\rangle}{#1\,\delimsize\vert\,\mathopen{}#2}
\usepackage{mathrsfs}
\usepackage{mathrsfs}
\usepackage[ qm]{qcircuit}
\usepackage{graphicx}
\usepackage{authblk}

\usepackage[sorting=none]{biblatex} 
\usepackage [english]{babel}
\usepackage [autostyle, english = american]{csquotes}
\MakeOuterQuote{"}

\usetikzlibrary{shapes.geometric}
\usetikzlibrary{decorations.pathmorphing}
\usetikzlibrary{shapes.arrows, fadings}

\input pdf-trans

\title{Deterministic Discrimination of Phase-Modified Permutation Oracles via Single Qubit Measurement}
\author[1,*]{Owen Root}
\affil[1]{Department of Physics and Astronomy, CUNY Hunter College, 695 Park Ave, New York, NY 10065, USA}

\affil[*]{oroot@gradcenter.cuny.edu}

\begin{document}

\maketitle

\begin{abstract}
I study a promise problem for an unknown unitary operator acting on an $n$-qubit system. The operator is promised to take one of two forms: either it implements a fixed permutation of computational basis states, or it implements the same permutation together with a conditional sign change determined by a designated input qubit. I show that these two cases can be distinguished with certainty using a single query to the unknown operator and a measurement of only one qubit. The procedure requires no ancilla qubits and uses only $n+1$ Hadamard gates in addition to the oracle call. The promise is intrinsically quantum, since the two cases differ only in their relative-phase structure and therefore have no direct classical counterpart in the usual black-box model.
\end{abstract}

\par
%
\section{Introduction}
Oracle identification and discrimination problems have been central to quantum computation since the seminal works of Deutsch \cite{Deutsch}, Deutsch and Jozsa \cite{Deutsch_Jozsa}, and Bernstein and Vazirani \cite{Bernstein_Vazirani}. These examples showed that superposition and interference can be used to extract global properties of a black-box operation with far fewer queries than are required classically. In this note I consider a different promise problem. Let $\hat{U}$ be a black-box unitary on an $n$-qubit register, promised to take one of two forms: either it permutes the computational basis states according to a fixed bijection, or it implements the same permutation together with a sign factor determined by a designated input qubit. The question is whether these two possibilities can be distinguished with certainty using a single query. I show that they can: an initial layer of Hadamard gates, one application of $\hat{U}$, a final Hadamard on the designated qubit, and a single-qubit measurement suffice, and no ancilla register is required. Unlike the Deutsch-Jozsa and Bernstein-Vazirani settings, the promise here concerns phase information alone, so the problem is best viewed as an intrinsically quantum oracle-discrimination task rather than as a classical decision problem admitting a quantum speedup.
\par
\section{Derivation}
Let us can label the two possible operators $\hat{U}=\hat{U}_1$ and $\hat{U}=\hat{U}_2$, respectively. 
\par
(Step 1) Consider the system in a general input state 
\begin{equation}
    \ket{\Psi}=\ket{i}=\bigotimes_{l=1}^n \ket{x_{il}}=\ket{x_{i1},x_{i2},\dots,x_{in}},
\end{equation}

where $i=0,1,2,\dots,2^n-1$ labels the states, $l=0,1,2,\dots,n$ labels the qubits, and $\ket{x_{il}}$ is the state of the $l$-th qubit in the total state $i$. Note that $\ket{\Psi}$ denotes the state vector of the system at a given step and that $\bigotimes$ denotes repeated tensor products. The effects of $\hat{U}_1$ and $\hat{U}_2$ can be expressed generally as 

\begin{equation}
    \hat{U}_1\ket{i}=\ket{k},
\end{equation}
\begin{equation}
    \hat{U}_2\ket{i}=f(x_{iL})\ket{k},
\end{equation}

where $f(x_{iL})$ is a function that is equal to 1 if the state of a particular qubit, the $L$-th qubit, $\ket{x_{iL}}$ is in the $\ket{0}$ state and equal to -1 if $\ket{x_{iL}}$ is in the $\ket{1}$ state. For example, if $n=2, L=1$, and the mapping from $\ket{i}\rightarrow\ket{j}$ is the identity mapping, $\hat{U}_1\ket{01}=\ket{01}$ but $\hat{U}_2\ket{01}=-\ket{01}$. Alternatively, $\hat{U}_1\ket{00}=\ket{00}$ and $\hat{U}_2\ket{00}=\ket{00}$.
\par
(Step 2) We then apply a Hadamard gate to each qubit in the system,

\begin{equation}
    \hat{H}^{\otimes n}\ket{\Psi}= 
    \frac{1}{\sqrt{2^n}}\sum_{j=0}^{2^n-1}(-1)^{i \cdot j}\ket{j}=
    \frac{1}{\sqrt{2^n}}\sum_{j=0}^{2^n-1}(-1)^{i \cdot j}\ket{x_{j1},x_{j2},\dots,x_{jn}},
\end{equation}

where $i\cdot j$ is the bitwise dot product of the (binary) $i$ and $j$ states. 
\par
Now we consider the different possibilities of the operator of $\hat{U}$ on the system. 
\par
(Step 3) If $\hat{U}=\hat{U}_1$ then 
\begin{equation}
    \hat{U}\ket{\Psi}=\hat{U}_1\ket{\Psi}=  
    \frac{1}{\sqrt{2^n}}\sum_{k=0}^{2^n-1}(-1)^{i \cdot j(k)}\ket{k}=
    \frac{1}{\sqrt{2^n}}\sum_{k=0}^{2^n-1}(-1)^{i \cdot j(k)}\ket{x_{k1},x_{k2},\dots,x_{kn}},
\end{equation}

where I have written the bitwise dot product as $i\cdot j(k)$ to emphasize that the sign of each $k$ term is dependent upon the state of the qubits in the $j$ state that each $k$ state is mapped from and not the state of the qubits in the $k$ state itself. 
\par

(Step 4) Now we apply a Hadamard gate to just qubit $L$ and leave the rest of the system alone, 

\begin{multline}
    \hat{H}_L\ket{\Psi}=\frac{1}{\sqrt{2^n}} \Bigg\{ \frac{1}{\sqrt{2}}\left(\ket{0}_L+\ket{1}_L\right)\left(\bra{0}_L \sum_{k=0}^{2^n-1}(-1)^{i \cdot j(k)}\ket{x_{k1},\dots,x_{kn}}\right)
    \\+
    \frac{1}{\sqrt{2}}\left(\ket{0}_L-\ket{1}_L\right)\left(\bra{1}_L \sum_{k=0}^{2^n-1}(-1)^{i \cdot j(k)}\ket{x_{k1},\dots,x_{kn}})\bigg\}\right),
\end{multline}

where the subscript $L$ on a bra or ket denotes that to be just the state of the $L$-th qubit, or more compactly as 

\begin{equation}
     \hat{H}_L\ket{\Psi}=\frac{1}{\sqrt{2^n}} \left( \frac{1}{\sqrt{2}}(\ket{0}_L+\ket{1}_L)a_1
    +
    \frac{1}{\sqrt{2}}(\ket{0}_L-\ket{1}_L)b_1\right),
\end{equation}

Notice that the signs of the terms in $a_1$ and $b_1$ depend upon the initial state of the $L$-th qubit, $\ket{x_{il}}$. If $\ket{x_{il}}=\ket{0}$ then $(-1)^{i\cdot j(k)}$ is the same for each term in in $a_1$ and $b_1$, that is $a_1=b_1$, while if $\ket{x_{il}}=\ket{1}$ then $(-1)^{i\cdot j(k)}$ gives the $k$-th term in $b_1$ the opposite sign as the $k$-th term in $a_1$, that is $a_1=-b_1$. This follows from states with $\ket{x_{jl}}=\ket{1}$ picking up a (-1) in step 2.
\par
Then, if $\ket{x_{iL}}=\ket{0}$, then the $\ket{1}_L$ terms cancel (due to the difference of signs in eq. (7)), leaving

\begin{equation}
    \ket{\Psi}=\frac{1}{\sqrt{2^n}}\left(\frac{1}{\sqrt{2}}2\ket{0}_La_1\right)=\frac{1}{\sqrt{2^{n-1}}}\ket{0}_La_1.
\end{equation}

But if $\ket{x_{iL}}=\ket{1}$, the $\ket{0}_L$ terms cancel leaving 

\begin{equation}
    \ket{\Psi}=\frac{1}{\sqrt{2^{n-1}}}\ket{1}_La_1.
\end{equation}
Therefore, (Step 5) if $\hat{U}=\hat{U}_1$, following this procedure, we will always find the $L$-th qubit in its initial state. 
\par
Now consider the case that $\hat{U}=\hat{U}_2$. Step 3 is then

\begin{equation}
    \hat{U}\ket{\Psi}=\hat{U}_2\ket{\Psi}=  
    \frac{1}{\sqrt{2^n}}\sum_{k=0}^{2^n-1}(-1)^{i \cdot j(k)} f(x_{j(k)L}) \ket{x_{k1},x_{k2},\dots,x_{kn}},
\end{equation}

where I have written the argument of $f$ as $x_{j(k)L}$ to emphasize that the value of $f$ for each $k$ term is dependent upon the state of the $L$-th qubit in the $j$ state that each $k$ state is mapped from and not the state of the $L$-th qubit in the $k$ state itself. 
\par
Step 4 is now 

\begin{multline}
    \hat{H}_L\ket{\Psi}=\frac{1}{\sqrt{2^n}} \Bigg\{ \frac{1}{\sqrt{2}}(\ket{0}_L+\ket{1}_L)\left(\bra{0}_L \sum_{k=0}^{2^n-1}(-1)^{i \cdot j(k)} f(x_{j(k)L}) \ket{x_{k1},\dots,x_{kn}}\right) 
    \\
    + \frac{1}{\sqrt{2}}(\ket{0}_L-\ket{1}_L)\left(\bra{1}_L \sum_{k=0}^{2^n-1}(-1)^{i \cdot j(k)} f(x_{j(k)L}) \ket{x_{k1},\dots,x_{kn}}\right) \Bigg\},
\end{multline},

or more compactly as 

\begin{equation}
     \hat{H}_L\ket{\Psi}=\frac{1}{\sqrt{2^n}} \left( \frac{1}{\sqrt{2}}(\ket{0}_L+\ket{1}_L)a_2
    +
    \frac{1}{\sqrt{2}}(\ket{0}_L-\ket{1}_L)b_2\right),
\end{equation}

Notice that here $a_2$ and $b_2$ have an additional contribution to the sign of $k$ terms, the factor $f(x_{j(k)L})$, compared to $a_1$ and $b_1$, which contributes a factor of $-1$ precisely to every state that also gets a factor of $-1$ from the $L$ qubits contribution to $(-1)^{i\cdot j(k)}$. So now, once again, we have a balanced or even situation in terms of the signs of the states in $a_2$ and $b_2$, that is, if $\ket{x_{iL}}=\ket{0}$, then $a_2=-b_2$ and if $\ket{x_{iL}}=\ket{1}$ then $a_2=b_2$. 
\par
Then, if $\ket{x_{iL}}=\ket{0}$, the $\ket{0}_L$ terms cancel, leaving

\begin{equation}
    \ket{\Psi}=\frac{1}{\sqrt{2^{n-1}}}\ket{1}_La_2.
\end{equation}

But if $\ket{x_{iL}}=\ket{1}$, the $\ket{1}_L$ terms cancel leaving 

\begin{equation}
    \ket{\Psi}=\frac{1}{\sqrt{2^{n-1}}}\ket{0}_La_2.
\end{equation}

Therefore (Step 5) if $\hat{U}=\hat{U}_2$, following this procedure, we will always find the $L$-th qubit in the opposite of its initial state. 
\par
\begin{figure}[h]
    \centering
    \scalebox{2.0}{
    \Qcircuit @C=1.0em @R=0.2em @!R { \\
    	 	\nghost{{q}_{0} :  } & \lstick{{q}_{0} :  } & \gate{\mathrm{H}} & \multigate{2}{\mathrm{U}}_<<<{0} & \qw & \qw & \qw & \qw\\
    	 	\nghost{{q}_{1} :  } & \lstick{{q}_{1} :  } & \gate{\mathrm{H}} & \ghost{\mathrm{U}}_<<<{1} & \qw & \qw & \qw & \qw\\
    	 	\nghost{{q}_{2} :  } & \lstick{{q}_{2} :  } & \gate{\mathrm{H}} \qw & \ghost{\mathrm{U}}_<<<{2} & \gate{\mathrm{H}} & \meter & \qw & \qw\\
    	 	\nghost{\mathrm{{c} :  }} & \lstick{\mathrm{{c} :  }} &  \cw & \cw & \cw & \dstick{_{_{\hspace{0.0em}}}} \cw \ar @{<=} [-1,0] & \cw & \cw\\
    \\ }}
    \caption{An example circuit implementation of the algorithm with $n=3$ and $L=2$.}
    \label{fig:example}
\end{figure}
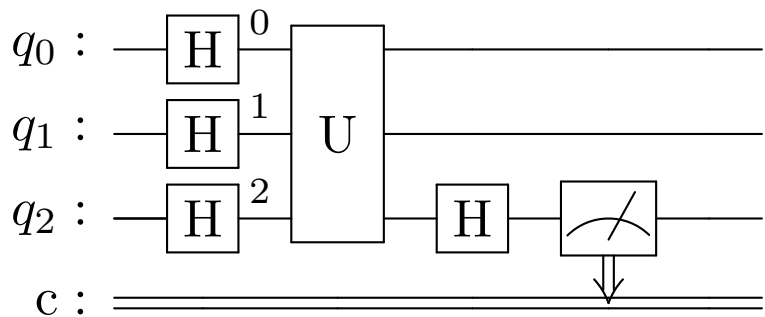

To summarize, the algorithm is as follows: (1) initialize the system in some known, unentangled state; (2) apply a Hadamard gate to each qubit, (3) act on the system with the unknown operator $\hat{U}$; (4) Apply a Hadamard gate to the $L$-th qubit; (5) measure the $L$-th qubit. If the $L$-th qubit is found to be in its initial state, we know that $\hat{U}=\hat{U}_1$ and if we find it flipped from its initial state, we know that $\hat{U}=\hat{U}_2$. See Fig. \ref{fig:example} for an example of an implementation of this algorithm. 

\section{Conclusion}

\par

Several remarks are worth making. First, the point of the result is modest. I do not claim practical utility, only that this promise problem admits exact discrimination using one oracle query and a final measurement on a single qubit. In that sense, the construction may be of conceptual interest as a minimal example of phase-sensitive oracle discrimination. Second, the problem is intrinsically quantum in character. The two oracles $\hat{U}_1$ and $\hat{U}_2$ induce the same permutation on computational basis labels and differ only through relative phases attached to amplitudes, so the distinction is not visible in a purely classical black-box model. For that reason, the present result should not be interpreted as a standard quantum-versus-classical query-complexity separation of the Deutsch-Jozsa or Simon type \cite{Simon}. Rather, it is an example of a promise problem whose formulation already depends on genuinely quantum structure. It would be interesting to determine whether this idea extends to broader families of phase-modified permutations, or whether related constructions could be useful in exact oracle discrimination, unitary certification, or other forms of quantum process characterization.

\printbibliography

\end{document}